\documentclass[preprint,superscriptaddress,preprintnumbers,aps,amsmath,amssymb]{revtex4}
\usepackage{graphicx} 
\usepackage{bm} 
\usepackage{longtable}
\usepackage{multirow}
\usepackage{color}
\usepackage{enumerate}

\newcommand{\bra}[1]{\langle #1 |}
\newcommand{\ket}[1]{| #1 \rangle}

\newcommand{\beq}{\begin{equation}}
\newcommand{\eeq}{\end{equation}}

\newcommand{\creation}[2]{\hat #1^{\dagger}_{#2}}
\newcommand{\annihilation}[2]{\hat #1_{#2}}

\newcommand{\be}{\begin{equation}}
\newcommand{\ee}{\end{equation}}
\newcommand{\bea}{\begin{eqnarray}}
\newcommand{\eea}{\end{eqnarray}}

\def\k{\mathbf{k}}

\def\ri{\mathbf{r}_i}

\def\xi{\mathbf{x}_i}

\begin{document}


\title{Investigating polaron transitions with polar molecules}

\author{Felipe Herrera }
\affiliation{Department of Chemistry, University of British Columbia,
  Vancouver, B.C., V6T 1Z1, Canada}
\affiliation{Department of Chemistry, Purdue University, West
  Lafayette, IN 47907, USA} 
\affiliation{Department of Chemistry and Chemical Biology,
	Harvard University, 12 Oxford St., Cambridge, MA 02138, USA}

\author{Kirk W. Madison}
\affiliation{Department of Physics and Astronomy, University of
  British Columbia, Vancouver, B.C., V6T 1Z1, Canada} 

\author{Roman V. Krems}
\affiliation{Department of Chemistry, University of British Columbia,
  Vancouver, B.C., V6T 1Z1, Canada}

\author{Mona Berciu} 
\affiliation{Department of Physics and Astronomy, University of
  British Columbia, Vancouver, B.C., V6T 1Z1, Canada}

\date{\today}

\begin{abstract}
We determine the phase diagram of a polaron model with mixed
breathing-mode and Su-Schrieffer-Heeger couplings and show that it
has two sharp transitions, in contrast to pure models which exhibit one (for Su-Schrieffer-Heeger coupling) or no (for breathing-mode coupling) transition. Our results indicate that the physics of realistic mixed polaron models is much richer than that of simplified models. We then show that ultracold molecules trapped in optical lattices can be used to study precisely this mixed Hamiltonian, and that the relative contributions of the two couplings can be tuned with external electric fields. The parameters of current experimental set-ups place them in the region where one of the transitions occurs. We propose a scheme to measure the polaron dispersion using stimulated Raman spectroscopy.
\end{abstract}


\maketitle

{\em Introduction:} Polarons, which are the low-energy dressed
quasiparticles appearing in the spectrum of particles coupled to
bosonic fields, have been of broad interest in physics ever since
their first study by Landau \cite{Landau}. The generic Hamiltonian for
a single polaron problem:
\begin{equation}
\label{i1}
{\cal H} = \sum_{k}^{} \epsilon_k c_k^\dagger c_k +
\sum_{q}^{} \hbar\Omega_q b^\dagger_q b_q + \sum_{k,q} g_{k,q}
c^\dagger_{k+q} c_k \left(b^\dagger_{-q} + b_q\right)
\end{equation}
includes the kinetic energy of the bare particle (first term), the
energy of the bosonic mode (second term) and their interaction (third
term), where the particle scatters by absorbing or emitting bosons.

There are two mechanisms for the particle-boson coupling, as the
presence of bosons can change (i) the potential or (ii) the kinetic
energy of the bare particle. The former always leads to a vertex $g_q$
independent of the particle's momentum $k$, while in the latter case
the vertex $g_{k,q}$ depends explicitly on both momenta. As an
example, consider interactions between electrons and
phonons. Vibrations of nearby atoms modulate
the potential energy of an electron. Much studied examples of such
type (i) interactions are the Holstein \cite{Holstein:1959}
and Fr\"ohlich \cite{Frohlich} models. At the same time, by modulating the
distance between lattice sites, lattice vibrations also affect the hopping
integrals. Such effects are described by type (ii) models like the
Su-Schrieffer-Heeger (SSH) model \cite{SSH:1988}.

Most of the early polaron studies focused on type (i) models, in
particular on the search for a self-trapping transition whereby at
strong coupling the bosons create a potential well so deep that it
traps the polaron. In the absence of impurities, however, this idea
has proved to be wrong: type (i) polarons have a finite
effective mass for any finite coupling strength. The absence of
transitions for type (i) models has been demonstrated analytically
\cite{Gerlach:1991}. These models have a smooth crossover
from light, highly mobile polarons at weak coupling to heavy, small
polarons at strong coupling.

This standard view of the polaron as a quasiparticle that becomes
heavier with increased coupling is now strongly challenged by results
for type (ii) models. Recent work has shown that in such models, the
polaron can be lighter than the bare particle, since the bosons affect
the particle's hopping so it may move more easily in their presence
\cite{Holger, Berciu:2010, Marchand:2010}.  The boson-mediated
dispersion can be quite different from that of the bare particle, for
example it may have a ground state with a different momentum
\cite{Berciu:2010}. If this happens, a sharp transition should occur
when the boson-mediated contribution to the dispersion becomes
dominant.  Indeed, such a transition was recently predicted for the
single SSH polaron \cite{Stojanovic:2008,Marchand:2010}.

These results are interesting not just for challenging a long-established paradigm, but also for raising the question of what happens in realistic systems, where a mix of both types of coupling is generally expected. For example, what is the fate of the transition in a mixed model if the coupling is varied smoothly from $g_{k,q}$ to
$g_q$?  This also makes it highly desirable to find systems described by such mixed Hamiltonians but where, unlike in solid-state systems, the various parameters can be tuned continuously so that the resulting polaron behaviour can be systematically investigated.

In this Letter we elucidate the evolution of this transition as the coupling interpolates between type (ii) SSH and type (i) breathing mode (BM) \cite{Goodvin:2008}.  Surprisingly, we find that the phase diagram has two sharp transitions, and that these may occur even when the type (i) coupling is dominant. This shows that polaron physics is much richer than generally assumed. We then show that this mixed Hamiltonian describes polar molecules trapped in an optical lattice, and moreover, that the parameters of current experimental set-ups place them in the region where a transition is expected to occur. Thus, experimental confirmation of these transitions is 
within reach.  Furthermore, we propose a detection scheme equivalent to Angle-Resolved Photoemission Spectroscopy (ARPES) in solid-state systems \cite{Andrea}, which directly measures the polaron dispersion and can therefore pinpoint the transition. Finally, as we discuss in conclusion, such experiments hold the promise not only to clarify many  other  effects of interplay between type (i) and type (ii) couplings on single polarons, but also to investigate the finite concentration regime to look for quantum phase transitions.

{\it Phase diagram:} Consider the particle-boson coupling 
\begin{equation}
 g_{k,q}= \frac{2i}{\sqrt{N}}\left\{\alpha\left[\sin(k+q)-\sin(k)\right]+ \beta\sin(q)\right\},
\end{equation}
where $N$ is the number of lattice sites in a one-dimensional chain. The $(k,q)$-dependent part, with energy scale $\alpha$, describes SSH coupling \cite{Marchand:2010} while the
$k$-independent part, with energy scale $\beta$, describes BM type modulations of the site energy by bosons \cite{Goodvin:2008}. We assume Einstein bosons $\Omega_q=\Omega$ and a free particle dispersion $\epsilon_k = + 2t \cos(k)$ with $t>0$ \cite{note2}.  Following Ref. \cite{Marchand:2010}, we define the effective SSH coupling strength $\lambda=2\alpha^2/(t\hbar\Omega)$ and the adiabaticity ratio $A = \hbar\Omega/t$. In addition, we introduce $R\equiv \beta/\alpha$ to characterize the relative strength of the two types of coupling. 

\begin{figure}[t]
\centering
\includegraphics[width=0.80\columnwidth]{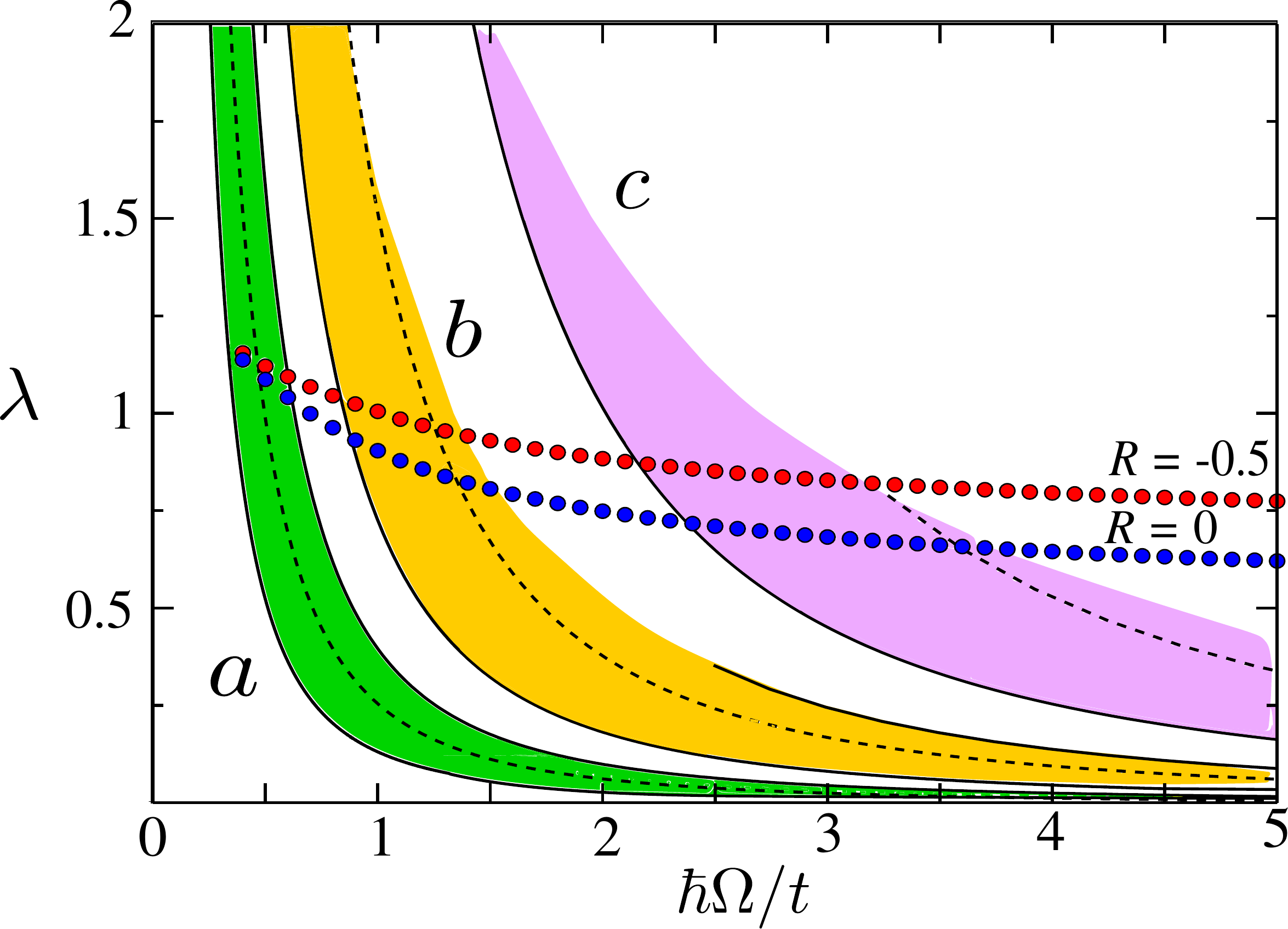}
\caption{(color online). Phase diagram $\lambda$ vs. $A$ at fixed
  $R$. Symbols show the location of the transition if $R=0$ (blue
  circles) and $R=-0.5$ (red circles). The three shaded regions
  represent molecules with large (a, LiCs), intermediate (b, RbCs) and
  small (c, KRb) dipoles, and dressing schemes 1 ($R=-0.5$ for a) and
  2 ($R=0$ for b and c), respectively. For each region, the three
  lines correspond to lattice constants of 256 nm (lower bound), 532
  nm (dashed line) and 775 nm (upper bound, not shown for region (c)
  because it is off-scale). For each curve $\hbar\Omega$ varies
  between 1 and 100 kHz.  }
\label{fig1}
\end{figure}

The SSH polaron ($R=0$) was predicted to undergo a sharp transition at a value $\lambda^*$ \cite{Marchand:2010}. Its physical origin is simple to understand in the anti-adiabatic limit $A \gg1$ where the SSH coupling leads to an effective next-nearest neighbor hopping $i\leftrightarrow i+2$ of the particle, by first creating and then removing a boson at site $i+1$ \cite{Marchand:2010}. Its amplitude is $t_2= -\alpha^2/(\hbar\Omega)=-\lambda t/2 <0$, so its contribution
$-2t_2\cos(2k)$ to the total dispersion has a minimum at $\pi/2$, unlike the bare dispersion which has a minimum at $\pi$.  If $4|t_2| = t$, corresponding to $\lambda^*={1\over 2}$ in the limit of $A\gg1$, a sharp transition marks the switch from a non-degenerate ground state with momentum $k_{gs}=\pi$ (for $\lambda < \lambda^*$) to a doubly-degenerate one with $| k_{gs}|\rightarrow {\pi\over2}$ (for $\lambda > \lambda^*$). As $A$ decreases the number of phonons in the polaron cloud increases. This renormalizes both hoppings $t\rightarrow t^*$, $t_2 \rightarrow t_2^*$, so $\lambda^*$ changes smoothly with $A$ as shown by the blue circles in Fig. \ref{fig1} (see also Fig. 4 of Ref. \cite{Marchand:2010}).  We also plot $\lambda^*$ for $R=-0.5$ (red circles), showing that the sharp polaron transition persists for coexisting type (i) and type (ii) couplings. These results were generated with the Momentum Average (MA) approximation, specifically its variational flavor where 
the polaron cloud is allowed to extend over any three consecutive sites \cite{Berciu:2010, Marchand:2010}.  For  $A\ge 0.3$, MA was shown to be  very accurate for both SSH and BM couplings \cite{Marchand:2010,Goodvin:2008}.

\begin{figure}[t]
\centering
\includegraphics[width=0.80\columnwidth]{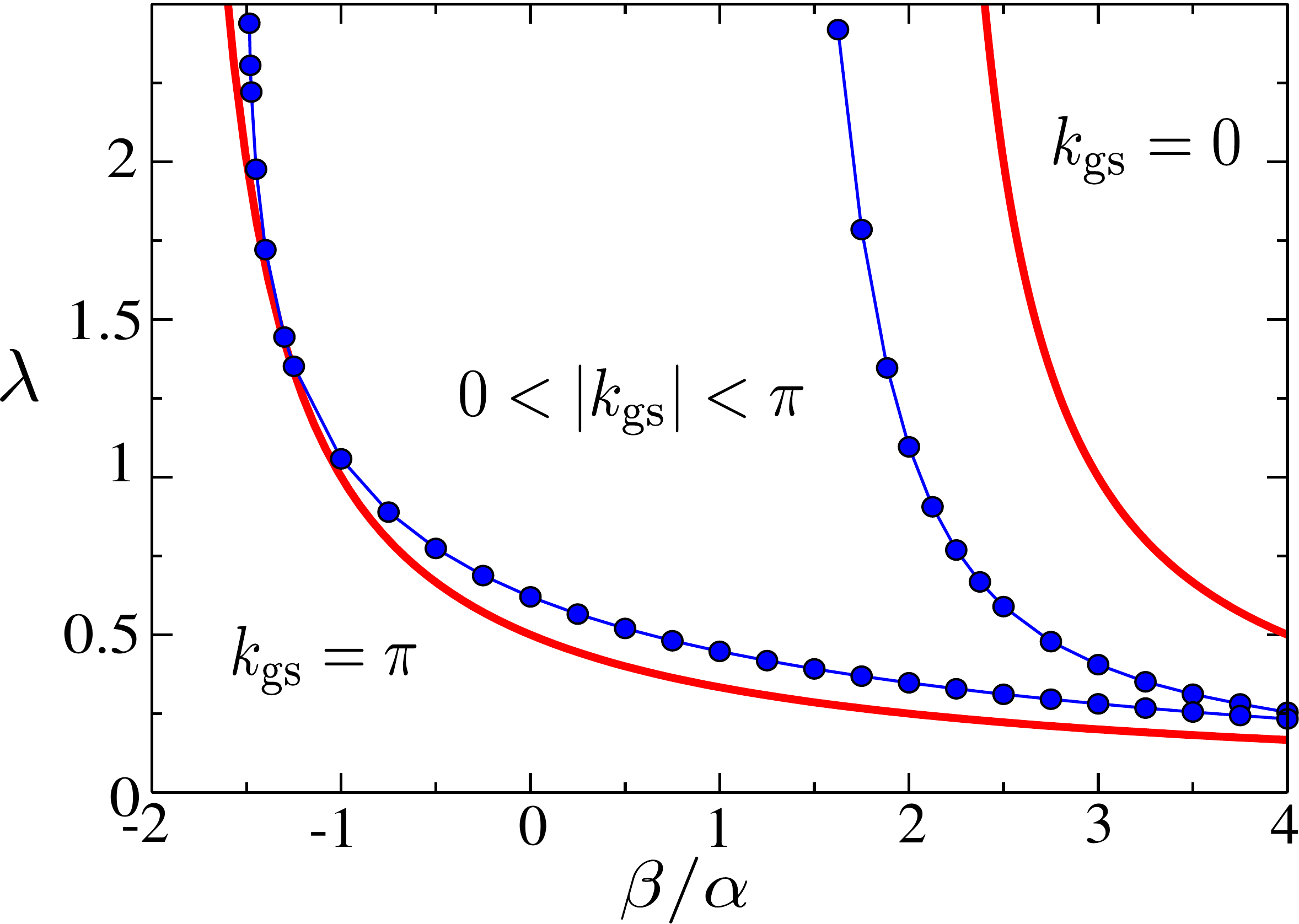} 
\caption{(color online). Phase diagram $\lambda$ vs. $R$ at fixed $A$, 
  showing two  sharp transitions: one from a non-degenerate ground state with
  $k_\text{gs}=\pi$ to a doubly-degenerate ground state with $0 <
  |k_\text{gs}|<\pi$, and the second back to a non-degenerate
  ground state with $k_\text{gs}=0$. The results are for
  $A=\hbar\Omega/t\rightarrow \infty$ (red lines)  and for $A=5$ (blue circles).}
\label{fig2}
\end{figure}

To understand the evolution of $\lambda^*$ with $R$, consider again the  limit $A \gg 1$. In addition to the second-nearest neighbor hopping $t_2$, there is now also a dynamically generated nearest-neighbor hopping $t_1 = 2 \alpha \beta/(\hbar \Omega)= R \lambda t$. This describes processes where the particle hops from site $i$ to $i+1$ leaving behind a boson at $i$ (SSH coupling) followed by absorption of the boson while the particle stays at $i+1$ (BM coupling); or vice versa, hence the factor of 2. The total nearest-neighbor hopping is thus $t^* = t- t_1$, and the transition now occurs when $ 4|t_2| = t^* \rightarrow \lambda^*= 1/( 2+R)$.  Thus, for $R <0$, the interference between the SSH and the BM couplings results in a larger effective $t^*$ leading to a larger $\lambda^*$. In particular, $\lambda^* \rightarrow \infty$ for $R \le -2$, i.e.  no transition occurs here. The lack of a transition is not surprising when  $R\rightarrow -\infty$, since here the BM coupling is dominant and pure $g_q$ models do not have transitions \cite{Gerlach:1991}. Our results show that for mixed SSH+BM coupling, the switch from having to not 
having a transition occurs abruptly at $R=-2$ if $A\gg 1$. This value must change continuously with $A$, therefore we expect this switch to always occur at a finite $R$.

This is confirmed in Fig. \ref{fig2}, where we plot $\lambda^*$ vs. $R$ for $A\rightarrow\infty$ and  $A=5$.
Surprisingly, we find  not just the transition at $\lambda^*\sim 1/(R+2)$, but also a second one which marks the crossing to a ground state with $k_{gs}=0$. Its origin is also easy to understand in the anti-adiabatic limit: if $R\lambda > 1$, $t^*$ is negative and favors a ground state at $k_\text{gs}=0$ instead of $k_\text{gs}=\pi$. For $A\gg 1$ this second transition is at $\lambda^*=1/(R-2)$ if $R>2$. At finite $A$, it moves towards smaller $(R,\lambda)$ values, see Fig. \ref{fig2}.

Interestingly, this shows that for $R\rightarrow + \infty$ there are two nearby transitions for the shift $k_{gs}=\pi$ to $k_\text{gs}=0$. This seems to contradict the proof that a type-(i) Hamiltonian cannot have transitions \cite{Gerlach:1991}, however, even for $R\rightarrow \infty$ this is a mixed Hamiltonian if $\lambda\ne 0$. The transition is indeed absent if $\alpha =0$.
This is an example of the rare occurrence where a perturbatively small term has a large effect on the behavior of the system.

{\it Cold molecule implementation:} Polar molecules in optical lattices can be used to implement Hamiltonian (\ref{i1}) in a wide region of the parameter space.  Specifically, we consider molecules prepared in the ro-vibrational ground state of the spinless electronic state $^1\Sigma$ and trapped on an optical lattice in the Mott insulator phase, as recently demonstrated experimentally \cite{Ospelkaus:2006,Chotia:2012}. We assume that there is at most one molecule per lattice site.

The dipole-dipole interaction between molecules in different sites can be modified by applying a DC electric field $\mathbf{E} = E_\text{DC}\hat{\mathbf z}$ \cite{Micheli:2006,Micheli:2007, Rabl:2007, Herrera:2010,Jesus:2010, Gorshkov:2011}.  Here we consider two schemes for dressing the rotational states of molecules with electric fields that are relevant for polaron observation, scheme 1 involving a DC electric field only, and scheme 2 involving combined optical and DC electric fields. For the former, we define the two-state subspace $\ket{g}=\ket{\tilde 0,0}$ and $\ket{e}=\ket{\tilde 1,0}$, where $\ket{\tilde N, M_N}$ denotes the field-dressed state that correlates adiabatically with the field-free rotational state $\ket{N,M_N}$. $N$ is the rotational angular momentum and $M_N$ is the projection of $N$ along the electric field vector. In this basis we define the pseudospin operator $\creation{c}{i}\equiv \ket{e_i}\bra{g_i}$ that creates a rotational excitation at site $i$. This excitation (the ``bare 
particle'') can be 
transferred between molecules in different lattice sites with an amplitude $t_{ij} = \gamma\, U_{ij}\,(1-3\cos^2\Theta)$, where $U_{ij} = d^2/|\mathbf{r}_i-\mathbf{r}_j|^3$, $d$ is the permanent dipole moment, $\ri$ is the position of molecule in site $i$, $\Theta$ is the polar angle of the intermolecular separation vector, $\gamma = \mu_{eg}^2/d^2\leq 1$ is the dimensionless transition dipole moment that depends on the strength of the DC electric field. The excitation hopping amplitude is finite even for vanishing field strengths. The field-induced dipole-dipole interaction shifts the energy of the state $\ket{e_i}$ by $D_i = \sum_j D_{ij}$. Here $D_{ij} = -\kappa\, U_{ij}\,(1-3\cos^2\Theta)$, where $\kappa = |\mu_g(\mu_e-\mu_g)|/d^2$ and $\mu_g$($\mu_e$) is the induced dipole of the ground(excited) state. Dipolar couplings outside this two-level subspace are suppressed when the electric field
separates state $\ket{e}$ from other excited states. 

The free quasiparticle dispersion is $\epsilon_k = \varepsilon_0 +2t\cos(k)$ where the site energy is $\varepsilon_0 = \hbar \omega_{eg}+D_{0}$, with the single-molecule rotational excitation energy $\hbar\omega_{eg}\sim 10 $ GHz and $t\equiv t_{12}$. The center-of-mass vibration of molecules in the optical lattice potential is coupled to their internal rotation through the dependence of $U_{ij}$ on $\mathbf{r}_i-\mathbf{r}_j$. For harmonic vibrations with linear coupling between internal and external degrees of freedom \cite{Rabl:2007, Herrera:2011}, the boson term in Eq. (\ref{i1}) describes lattice phonons whose spectrum depends on the trapping  laser intensity and the DC electric field \cite{Herrera:2011}. Here we consider weak DC fields and moderate trapping frequencies which give a gapped and nearly dispersionless (Einstein) phonon spectrum with frequency $\Omega$.

With these definitions, the system is described by SSH and BM-like couplings with the energy scales $\alpha = -3(t_{12}/a_L)\sqrt{\hbar/2m\Omega}$ and $\beta = -3(D_{12}/a_L)\sqrt{\hbar/2m\Omega}$, respectively, where $m$ is the mass of the molecule and $a_L$ is the lattice constant. The ratio $R = \beta/\alpha=\mu_g(\mu_e-\mu_g)/\mu_{eg}^2$ is independent of the intensity of the trapping laser or of the orientation of the array with respect to the DC field. In the field dressing scheme 1 we have $|R| < 1/2$ for $dE_{\text{DC}}/B_e\leq1$. 
In the combined AC-DC dressing scheme 2, the same DC field strength and orientation is used as above, but an additional two-color Raman coupling redefines the two-level subspace as $\ket{g}=\sqrt{a}\;\ket{\tilde 0,0}+\sqrt{1-a}\;\ket{\tilde 2,0}$ and $\ket{e}=\sqrt{b}\;\ket{\tilde 1,0}+\sqrt{1-b}\;\ket{\tilde 3,0}$ (details in the Supplementary Information). This dressing scheme can be used to effectively enhance the hopping amplitude by a factor $f>1$ yielding $\alpha\rightarrow f\alpha$, without changing the value of $\beta$, nor the phonon dispersion. When using this dressing scheme, any point in the phase diagram transforms as $\lambda\rightarrow f\lambda$ and $A\rightarrow A/f$, thus shifting the system towards stronger SSH couplings.

The frequency of lattice phonons in a 1D array is $\Omega = (2/\hbar)\sqrt{V_0E_R}$ where $V_0$ is the lattice depth and $E_R=\hbar^2\pi^2/2ma_L^2$ is the recoil energy. The particle-boson coupling can thus be written as $\lambda=18 (E_R/t)(\pi  A)^{-2}$. The shaded regions in Fig. \ref{fig1} show accessible points in the polaron phase diagram $(\lambda, A)$ for LiCs, RbCs and KRb molecules, illustrating the flexibility in varying the Hamiltonian parameters when using the two field dressing scenarios and different experimental settings.
Figure \ref{fig1} shows that the transition characterized by the shift from a non-degenerate ground state $k_\text{gs}=\pi$ to a degenerate ground state $0<|k_\text{gs}|<\pi$ can be studied using molecular species with moderate dipole moments such as RbCs, in lattices with a site separation $a_L\approx 500$ nm. However, the transition is easier to observe for molecules with large dipole moments such as LiRb and LiCs. For weakly dipolar molecules such as KRb, the strong coupling region can be achieved using $a_L<500$ nm. 

\begin{figure}[t]
\centering
\includegraphics[width=0.80\textwidth]{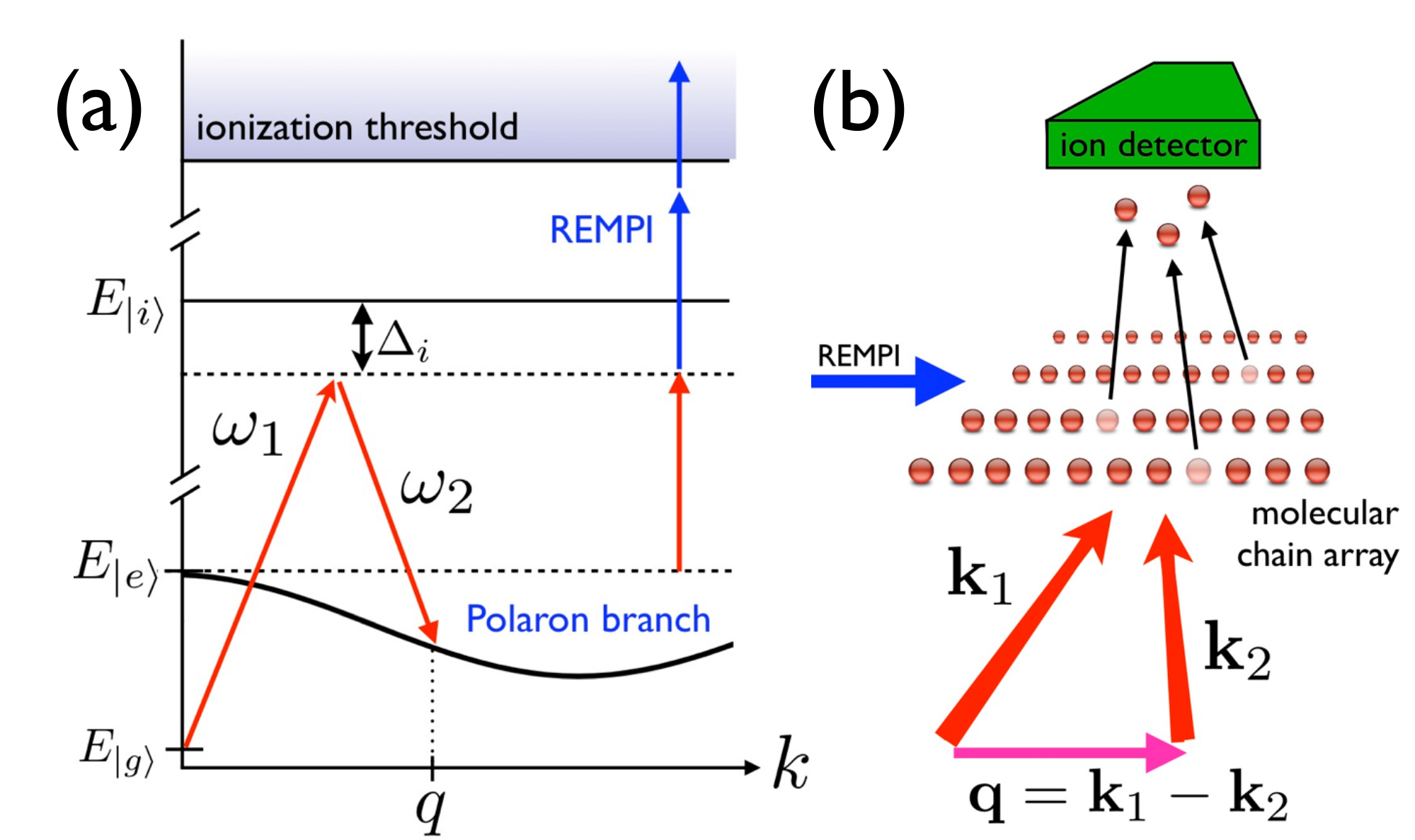}
\caption{(Color online) (a) A two-photon stimulated Raman transition
  creates a polaron state with a well defined momentum
  $q$ and energy $\omega = \omega_1-\omega_2$. (b) The presence
  of the quasiparticle is subsequently detected using resonantly-enhanced multi-photon ionization
  (REMPI). }
\label{fig4}
\end{figure}

The most direct way to detect the transition is to measure the polaron dispersion. We propose the stimulated Raman spectroscopic scheme illustrated in Fig. \ref{fig4} to achieve this goal. 
We consider a one-dimensional array initially prepared in the absolute ground state $\ket{g}=\ket{g_1,\ldots,g_N}\ket{\{0\}}$, where
$\ket{\{0\}}$ is the phonon vacuum.  We consider two linearly-polarized laser beams with wavevectors arranged such that $\k_1-\k_2$ is parallel to the molecular array. If the laser beams are far-detuned from any vibronic resonance the effective light-matter interaction
operator can be written as $ \hat V(t) = -g_N [ \creation{c}{q} \text{e}^{-i\omega t} + \annihilation{c}{q}\text{e}^{i\omega t}]$, where $g_N$ is a size-dependent coupling energy proportional to the amplitudes of both laser beams, $q=|\k_1-\k_2|$ and $\omega =\omega_1-\omega_2$ are the net momentum and energy transferred from the fields to the molecules. For short interaction times (linear response), the system is excited from $\ket{g}$ into the one-particle sector with a probability proportional to the spectral function $\mathcal{A}(q,\omega)=-\text{Im}[G(q,\omega)]/\pi$ where $G(q,\omega)=\langle g|\annihilation{c}{q}(\hbar \omega - {\cal H} + i \eta)^{-1}\creation{c}{q}|g\rangle$ is the one-particle Green's function. As a result, for any $q$ the polaron energy $E_q$ equals the energy $\hbar\omega$ of the lowest-energy peak in $\mathcal{A}(q,\omega)$, in analogy with ARPES measurements \cite{Andrea}.

To measure $\mathcal{A}(q,\omega)$, the stimulated Raman excitation rate can be determined using state selective resonance enhanced multi-photon ionization (REMPI), see Fig. \ref{fig4}(b). With some probability this converts the rotational excitation (the ``particle'') into a molecular ion which can be extracted from the chain and detected by a multi-channel plate ion detector. Ionization and subsequent detection efficiencies for a 2 step REMPI processes can easily exceed 20\% \cite{Stwalley:2011}, and a properly gated integrator can resolve the arrival of a single molecular ion. Using a 3D lattice, a set of uncoupled parallel 1D arrays can be realized and excited simultaneously, increasing the signal to noise ratio of the detection step. 

In summary, we presented the first (to our knowledge) phase diagram for a mixed type polaron Hamitonian \cite{note3}, which showed that
polaron physics is much richer than previously thought and that sharp transitions may occur even for dominantly type-(i) Hamiltonians ($R\gg 1$). We showed that polar molecules trapped in optical lattices can be used to study this physics, and proposed an ARPES-like detection scheme to directly measure the polaron dispersion and thus identify the transitions expected to occur in such systems. 

Polarons with type (i) coupling have been observed for a single atomic impurity immersed in a Fermi gas \cite{Schirotzek:2009,Nascimbene:2009,Will:2011}, and proposed for realization in lattice setups using atom-molecule systems \cite{Ortner:2009}, self-assembled crystals in strong DC fields \cite{Rabl:2007}, and recently trapped ions \cite{Stojanovic:2012}. Realization of type (ii) coupling has been considered using Rydberg atoms \cite{Hague:2012}. Using these systems it would be possible to explore the region of the phase diagram with $k_\text{gs}=0$ for $t<0$ (or equivalently $k_\text{gs}=\pi$ for $t>0$). Here we showed that using trapped molecules in weak DC electric fields it is possible to explore transitions into the phase with $0<|k_\text{gs}|<\pi$, where novel polaron physics is expected to occur \cite{Marchand:2010,Stojanovic:2008}.



Many other aspects of single polaron physics can be investigated with trapped polar molecules. Examples include investigating the effects of dispersive
phonons (most theoretical work assumes Einstein bosons), or novel effects resulting from quadratic particle-boson coupling in the strong coupling regime.
Studying polaron phase diagrams in higher dimensions is easily achieved with the same experimental scheme. Generalizations to studies of bi-polarons are also of significant interest, to understand the pairing mechanism for dominantly type-(ii) models (most bi-polaron studies are for type-(i) Holstein and Fr\"ohlich models). Finally, one may also be able to adapt the polar molecules systems to study finite polaron concentrations and search for quantum phase transitions \cite{QPT}.

{\it Acknowledgements:} Work supported by NSERC and CIFAR. FH would also like to thank NSF CCI center ``Quantum Information for Quantum Chemistry
(QIQC)'', Award number CHE-1037992.

\clearpage

\begin{center}
 \section*{\Large {Supplementary Material}}
\end{center}

\subsection*{Rotational state dressing schemes}

\noindent
{\it Static field dressing}.- 
This dressing scheme involves an homogeneous DC electric field acting on a one-dimensional array of $\mathcal{N}$ polar molecules. The field can have an arbitrary orientation with respect to the array axis, described by the polar angle $\Theta$. We consider $^1\Sigma$ molecules each prepared in its lower vibrational and electronic state. The rotational state of the molecules is described by the pendular state $\ket{\tilde N,M}$, which is an eigenstate of $\hat N_z$ and $\hat H_\text{R} = B_e \hat N^2 - dE_\text{DC}\cos\theta$, where $\hat N$ is the rotational angular momentum operator, $B_e$ is the rotational constant, $d$ is magnitude of the permanent dipole moment, and $\theta$ is the polar angle of the internuclear axis \cite{Carrington:2003}. 
We use the two-level subspace $\ket{g}=\ket{\tilde 0,0}$ and $\ket{e}=\ket{\tilde 1,0}$ to define the particle creation operator $\creation{c}{_i}=\ket{e_i}\bra{g_i}$. If the state $\ket{\tilde 1,0}$ is energetically separated from the degenerate states $\ket{\tilde 1, \pm 1}$ by an amount larger than the dipole-dipole interaction energy between neighbouring molecules in an array, the two-level approximation is valid and the dipole-dipole interaction operator reduces to $\hat V_\text{dd} = |\mathbf{r}_{ij}|^{-3} \left(1-3\cos^2\Theta\right)\hat d_0(i)\hat d_0(j)$, where $\mathbf{r}_{ij}$ is the intermolecular separation vector, $\Theta$ its polar angle with respect to the dc field, and $\hat d_p$ is the $p$-component of the space-fixed dipole operator in spherical coordinates. This simplification of the dipole operator is accurate for electric field strengths $E_\text{DC}>1$ V/cm, regardless of its orientation with respect to the molecular array. The dipolar energies $t_{12}=\bra{e_ig_j}\hat V_\text{dd}\ket{
g_ie_j}$ and $D_{12} = \bra{e_1g_2}\hat V_\text{dd}\ket{e_1g_2}-\bra{g_1g_2}\hat V_\text{dd}\ket{g_1g_2}$ are evaluated by diagonalizing $\hat H_\text{R}$ numerically and using the eigenvectors to compute the matrix elements of $\hat V_\text{dd}$ for a given value of the field strength $E_\text{DC}$.

{\it Combined static and infrared dressing}.- This schme involves a DC electric field as before and an additional infrared laser fields that induce Raman couplings between rotational states of the same parity. The Raman coupling scheme is illustrated in Fig. 1. It consists of two uncoupled lambda systems $\Lambda_1=\{\ket{^1\Sigma,v=0}\ket{\tilde 0,0}, \ket{^1\Sigma,v=1}\ket{\tilde 1,0},\ket{^1\Sigma,v=0}\ket{\tilde 2,0}\}$ and $\Lambda_2 =\{\ket{^1\Sigma,v=0}\ket{\tilde 1,0},\ket{^1\Sigma,v=1}\ket{\tilde 2,0},\ket{^1\Sigma,v=0}\ket{\tilde 3,0}\}$. Here $\ket{^1\Sigma,v}$ denotes the electronic and vibrational quantum numbers. Four linearly polarized infrared fields are needed to implement the scheme. The corresponding transitions are far detuned by $\delta\omega>10^2$ MHz from other rotational lines in the presence of a DC field $E_\text{DC}\sim 1$ kV/cm.

Let $\Omega = |\Omega|\text{e}^{i\phi}$ be the Rabi frequency associated with a dipole-allowed transition. For each lambda system the adiabatic state with zero eigenvalue (dark state) can be written as [2] 
\begin{equation}
\ket{\lambda_0}=\cos\alpha(t)\ket{\epsilon_1}-\sin\alpha(t)\exp(i\beta)\ket{\epsilon_3},
\end{equation}
where the mixing angle $\alpha(t)$ is defined by $\tan\alpha(t)=|\Omega_P(t)/\Omega_S(t)|$, and the relative phase of the fields is $\beta = \phi_P-\phi_S$. The RWA holds when $\text{max}\left\{\Delta,\Omega_P,\Omega_S\right\}\ll B_e$. The two lambda systems in Fig. 1 are uncoupled due to the anharmonicity of the rotational spectrum for small detunings $\Delta_1$ and $\Delta_2$. Selection rules restrict the coupling between rovibrational states with different angular momentum projection $M$.

\begin{figure}[t]
 \includegraphics[width=0.90\textwidth]{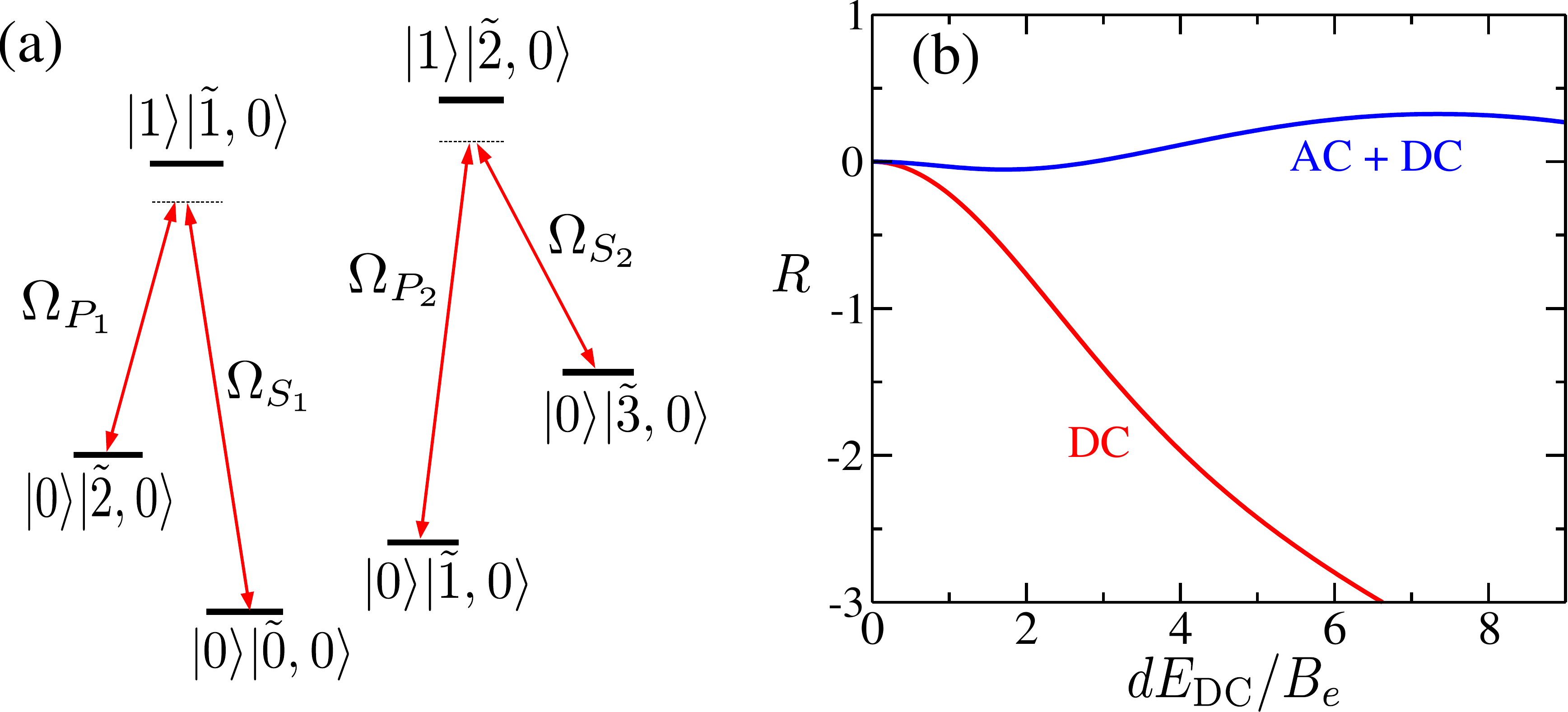}
\caption{(a) Infrared dressing of rotational states in the presence of a dc electric field. Two independent lambda systems are used to construct the dark state superpositions $\ket{g}=\sqrt{a}\ket{\tilde 0,0}+\sqrt{1-a}\ket{\tilde 2,0}$ and $\ket{e}=\sqrt{b}\ket{\tilde 1,0}+\sqrt{1-b}\ket{\tilde 3,0}$. (b) Dipole-dipole energy ratio $R = D_{12}/t_{12}$ as a function of the DC fields strength $E_\text{DC}$ for DC field rotational dressing ($a=b=1$) and combined AC-DC dressing with $a=b=1/2$.}
\end{figure}

We want to use the dark states of the lambda systems $\Lambda_1$ and $\Lambda_2$ to define the two-level subspace $\ket{g}$ and $\ket{e}$, respectively. It is in this field-dressed space where excitation hopping processes take place in a molecular array.
Let the molecules be initially prepared in the rovibrational ground state $\ket{g}=\ket{^1\Sigma,v=0}\ket{\tilde 0,0}$. We assume that only the Stokes fields $\Omega_{S_1}(t)$ and $\Omega_{S_2}(t)$ are present. In this case the mixing angles $\alpha_1(t)$ and $\alpha_2(t)$ vanish and the dark states are the same as the DC dressing scheme $\ket{g}=\ket{^1\Sigma,v=0}\ket{\tilde
0,0}$ and $\ket{e}=\ket{^1\Sigma,v=0}\ket{\tilde 1,0}$. By slowly turning on the pump fields $\Omega_{P_1}(t)$ and $\Omega_{P_2}(t)$ the dark states become  
\begin{eqnarray}
 \ket{g} &=& \cos\alpha_1(t)\ket{^1\Sigma,v=0}\ket{\tilde 0,0}+\sin\alpha_1(t)\ket{^1\Sigma,v=0}\ket{\tilde 2,0}\nonumber,\\
\ket{e} &=& \cos\alpha_2(t)\ket{^1\Sigma,v=0}\ket{\tilde 1,0}+\sin\alpha_2(t)\ket{^1\Sigma,v=0}\ket{\tilde 3,0}\nonumber.
\end{eqnarray}
The Raman lasers have the relative phases $\beta_1=\beta_2 = \pi$, which can be achieved experimentally. The states $\ket{g}$ and $\ket{e}$ can be manipulated independently by adiabatically tuning the pump laser intensities relative to the Stokes beams. We consider the specific situation where $\alpha_1 =\alpha_2= \pi/2$. Omitting the labels $v$ and $M$, the excitation hopping amplitude $t_{ij}$ is then given by
\begin{eqnarray}
 t_{ij} &\approx&\frac{1-3\cos^2\Theta}{|\mathbf{r}_{ij}|^3}\left(\frac{1}{4}\right)\left\{ d_{10}^2+d_{21}^2 + d_{32}^2 +2d_{21}d_{10} +2d_{32}d_{10}+2d_{21}d_{32}\right\}
\label{eq:type II t12}
\end{eqnarray}
where $d_{NN'}=\bra{\tilde N}\hat d_{0}\ket{\tilde N'}=d_{N'N}$. We have ignored couplings with $\Delta \tilde N>1$, which are suppressed in the presence of weak dc electric fields $E_\text{DC}\leq B_e/d$. Equation \ref{eq:type II t12} becomes exact in the limit $E_\text{DC}\rightarrow 0$ where the transition dipole moments can be evaluated analytically using angular momentum algebra [1].
The transition dipole moments in Eq. (\ref{eq:type II t12}) have the same sign and therefore contribute to the enhancement of the hopping amplitude $t_{12}'=f\,t_{12}$ with respect to the pure dc dressing scheme. The diagonal energy $D_{12}$ has a similar behaviour as in the type-I coupling only for weak dc electric fields $dE_\text{DC}/B_e\ll 1$. For non-perturbative fields, contributions from higher rotational states $\tilde N\geq 2$ modify significantly its static field dependence. 
In Fig. 1(b) we show the achievable values of the ratio $R=D_{12}/t_{12}$ using the combined AC-DC dressing scheme described above, as function of the DC field strength. The hopping amplitude is enhanced by a factor $f\approx 2$ with respect to the static scheme. The ratio $R$ has the same sign as the static dressing case ($R<0$) for weak fields $E_\text{DC}\leq 2 B_e/d$, but is up to an order of magnitude smaller. The ratio changes sign at $E_\text{DC}\approx 2.8 B_e/d$ and has a local maximum $R_\text{max}\approx 0.32$ at $E_\text{DC}=7B_e/d$.


{\it Phonon dispersion.-} In this work the dispersion of the phonon spectrum is ignored due to the current limitations of the Momentum Average (MA) approximation we use in the main text to solve the polaron problem [3]. 
For a 1D array the phonon frequency can be written as $\omega_q = \omega_0\sqrt{1+12\rho\gamma(q)}$, where $\rho=V^{gg}_{12}/V_0$ is the ratio between the ground state dipolar energy $V_{12}^{gg}=\bra{g_1g_2}\hat V_\text{dd}\ket{g_1g_2}$ and the optical lattice depth $V_0$, and $\gamma(q)=2\sin^2(q)$ [4]. 
The dispersion is determined by the parameter $\rho$, which can be kept small for shallow lattices $V_0\approx 18-20 E_R$ in the static dressing scheme only for weak dc fields $E_\text{DC}\leq B_e/d$. For higher fields we have $|V_{12}|^{gg}>|t_{12}|$, thus the phonon dispersion becomes an important energy scale. For combined AC-DC dressing, the phonon dispersion can be safely ignored since $|V^{gg}_{12}|\ll |t_{12}|$ over the entire range of electric fields 
in Fig. 1. 



\vspace{40pt}

\begin{enumerate}[(1)]
\item J. Brown and A. Carrington, {\it Rotational Spectroscopy of Diatomic Molecules}, (Cambridge University Press, 2003).
\item K. Bergmann, H. Theuer and B. W. Shore, Rev. Mod. Phys. {\bf 70}, 1003 (1998).
\item G. L. Goodvin and M. Berciu, Phys. Rev. B {\bf 23}, 235120 (2008).
\item F. Herrera and R. V. Krems, Phys Rev. A. {\bf 84}, 051401(R) (2011).
\end{enumerate}


\end{document}